# THE PLANCK NUMBERS AND THE ESSENCE OF GRAVITATION: PHENOMENOLOGY

Timashev Serge

Karpov Institute of Physical Chemistry

We introduce phenomenological understanding of the electromagnetic component of the physical vacuum, the EM vacuum, as a basic medium for all masses of the expanding Universe, and "Casimir polarization" of this medium arising in the vicinity of any material object in the Universe as a result of conjugation of the electric field components of the EM vacuum on both sides ("external" and "internal") of atomic nucleus boundary of the each mass with vacuum. It is shown that the gravitational attraction of two material objects in accordance with Newton's law of gravity arises as a result of overlapping of the domains of the EM vacuum Casimir polarization created by atomic nuclei of the objects, taking into account the long-range gravitational influence of all masses of the Universe on each nucleus of these objects (Mach's idea). Newton's law of gravitational attraction between two bodies is generalized to the case of gravitational interaction of a system of bodies when the center of mass of the pair of bodies shifted relative to the center of mass of the system. The unique smallness of gravitational interactions as compared with the fundamental nuclear (strong and weak) and electromagnetic ones is determined by the ratio of the characteristic size of the domain of EM vacuum Casimir polarization in the vicinity of atomic nuclei to the Hubble radius of the Universe.

1. INTRODUCTION

In 1899 M. Planck, in a purely arithmetic fashion, from the dimensional considerations, introduced the parameters of length $a_{Pl}$, time $t_{Pl}$ and mass $m_{Pl}$ by way of combining the basic universal constants $\hbar$, $c$ and $G$: Planck's constant, the velocity of light in vacuum and the gravitational constant, respectively [1]:

$$a_{Pl} = 2^{3/4}\sqrt{\frac{G\hbar}{c^3}} = 2^{1/2}\frac{\hbar}{m_{Pl}c} \approx 2.64 \cdot 10^{-33} cm; \quad t_{Pl} = \frac{a_{Pl}}{c} = 2^{3/4}\sqrt{\frac{G\hbar}{c^5}} \approx 0.88 \cdot 10^{-43} s;$$

$$m_{Pl} = 2^{-1/4}\sqrt{\frac{\hbar c}{G}} \approx 1.78 \cdot 10^{-5} \ g \qquad (1)$$

(the choice of numerical coefficients in (1) is clarified in [2, 3]). It should be noted that the values of the parameters of length $a_{Pl}$ and time $t_{Pl}$ are absolutely unattainable in the conditions of physical experiment in all times. Also unattainable on the accelerators is the energy $E_{Pl} = m_{Pl}c^2$, defined by the Planck mass, though in the value of the mass $m_{Pl}$ itself there is nothing enigmatic. To give the entire set of the Planck numbers an external unity of "unattainability" in experiments, one can consider, instead of the parameter $m_{Pl}$, the Planck power $w_{Pl}$, which can be represented in the form [3]:

$$w_{Pl} = \frac{m_{Pl}c^2}{t_{Pl}} = \frac{c^5}{2G} \approx 1.8 \cdot 10^{59} \ erg/s. \qquad (2)$$

It is known [4] that the Planck numbers have become a kind of heuristic reference point in solving a long-standing problem: establishing the essence of the gravitational interactions and constructing a theory of quantum gravity. In 1964, in the lectures at Cornell University on the nature of fundamental interactions, R. Feynman emphasized that "… up to today, from the time of Newton, no one has invented another theoretical description of the mathematical machinery behind this law which does not either say the same thing over again, or make the mathematics harder, or predict some wrong phenomena. So there is no model of the theory of gravitation today, other than the mathematical form" ([5], p. 39). And this was said despite the fact that Einstein's theory of gravitation, the General Relativity Theory (GRT), had already demonstrated its capabilities for understanding the Universe. The first predicted and verified experimental corollaries of GRT were three effects: an additional shift of the perihelion of Mercury's orbit in comparison with the predictions of Newton's mechanics; deflection of the light ray in the gravitational field of the Sun; gravitational red shift or time dilation in the gravitational field [7]. According to GRT, the bodies attract not because gravitation as a certain force acts on them, but as a consequence of the fact that the masses in the space form a curved four-dimensional space-time, whereas the particles move along the "geodesic lines", the world lines of physical bodies providing the shortest, "fastest" paths. However, the "internal mechanism" of the gravitational interaction, the first causes of formation of such geometrically curved paths and especially the reasons of smallness of gravitational forces, which are weaker by almost 40 orders of magnitude than the electromagnetic interactions, remain unknown in contemporary physics. Here we should note that, creating GRT (1921), Einstein, being aware of the limitation of the available information for solving this kind of problems, wrote: "We have seen, indeed, that in a more complete analysis the energy tensor can be regarded only as a provisional means of representing matter. In reality, matter consists of electrically charged particles, and is to be regarded itself as a part, in fact, the principal part, of the electromagnetic field. It is only the circumstance that we have not sufficient knowledge of the electromagnetic field of concentrated charges that compels us, provisionally, to leave undetermined in presenting the theory, the true form of this tensor" [6, p. 87].

In fact, the last statement means that the question of whether the gravitational constant $G$ is a carrier of information about gravity as a fundamental interaction and can be regarded as a universal constant together with $\hbar$ and $c$ remained open for Einstein and still remains open even after nearly 100 years since the creation of the GTR. Therefore, the collection of the three basic Planck numbers is *a priori* insufficient to become even a heuristic guideline for the realization of the well-known $\hbar Gc$-plan, which began to be developed by M. Bronshtein already in 1933 [4], and which meant to achieve an epistemological unity of the quantum theory (with its constant $\hbar$), special relativity theory (with its constant $c$) and the theory of gravitation (with its constant $G$).

Since the theoretical science is not yet able to give answers to Feynman's questions about the nature of gravity, a starting position in the search of new ideas for the "revealing the reality exactly how it shows itself *before* science turns to it with its questions" ([8], p.8) can be phenomenology ("phainomenon" is the manifesting; "logos" is a concept, doctrine). Phenomenology can be considered as a general methodology of cognizing the essence of a phenomenon, if one follows the basic maxim of Husserl's realist phenomenology "Back to the things themselves!" without "any attempts of premature systematization", to "studying 'the things themselves', not allowing ... any abuse of the givenness" [9]. Moreover, the analysis of the experimentally obtained information should be based on purely philosophical tradition with the a priori use of transcendental images which are outside the direct experimental tests, but reflect the main essence of the phenomenon under study, such as the "thing in itself" (*das Ding an sich*) and "now", just as in the classical mechanics of Newton there were introduced the images of "single mass point" and "inertial reference system". As the experience of developing the methodology of retrieving information from complex signals shows [10], such a view on phenomenology, while following the indicated philosophical traditions, allows not only focusing on achieving a purely pragmatic goals in solving technical problems and believing that in the framework of phenomenology "the cognition of the laws of nature is to a great extent even excessive" [11], but also setting the goals of penetration into the physical essence of the phenomenon under investigation, establishing the basic postulates, which can be a basis for the subsequent development of the corresponding theoretical models. It is precisely this approach to establishing the nature of gravity − by introducing a number of new transcendent entities – that we propose in this paper (see also [12, 13]).

2. PHENOMENOLOGICAL RELATIONS FOR THE BASIC UNIVERSAL CONSTANTS AND THE INTERACTION CONSTANTS

Since the gravitational constant $G$ cannot be regarded as a universal constant, to understand the physical genesis of this constant, and hence to establish the physical nature of the Planck numbers (1) and (2), it is necessary to extend the collection of universal constants, in addition to $\hbar$ and $c$, through which the quantity $G$ can be represented. To this end, let us use a result of S. Weinberg [14], who noticed the approximate equality:

$$\hbar \approx \frac{1}{2\pi} G^{1/2} m_\pi^{3/2} R_H^{1/2}, \qquad (3)$$

where $R_H$ is the characteristic ("Hubble") radius of the Universe ($R_H \sim 1.3 \cdot 10^{28}$ cm [15]) and $m_\pi$ is the mass of $\pi$-meson ($m_\pi \approx 140$ MeV/$c^2$), and represent the expression for $\hbar$ in the form of the equality [2, 13]:

$$\hbar = \frac{1}{2\pi} G^{1/2} m_Q^{3/2} R_H^{1/2}. \qquad (4)$$

Here we introduce a new energy parameter $E_Q = m_Q c^2 \approx 209.5$ MeV, the value of which is defined in such a way that there holds the well-known connection between the de Broglie wavelength and the particle momentum. The introduced parameter $E_Q$ can be considered as a specific energy of reorganization of the physical vacuum, corresponding to the elementary quantum of action. This value turns out to be corresponding to the energy scale of 200 MeV, considered in quantum chromodynamics [16]: at the corresponding intranuclear temperatures there occurs a phase transition in the nuclear matter, quarks are no longer bound in the nucleons, and the quark-gluon plasma is formed.

The expression (4) can be conveniently represented in the form:

$$G = \frac{(2\pi\hbar)^2 H}{m_Q^3 c} = \frac{2\pi^2 cH}{m_Q} a_Q^2 = 2^{3/2} \pi^2 \frac{\hbar c}{m_Q^2} \frac{a_Q}{R_H}. \qquad (5)$$

Here $a_Q = 2^{1/2} \hbar / m_Q c \approx 1.3 \cdot 10^{-13}$ cm is the "Bohr radius" (see below), corresponding to the mass $m_Q$ [2]. We will give to the quantity $m_Q = E_Q/c^2 \approx 3.72 \cdot 10^{-25}$ g the meaning of a fundamental one, characteristic for the strong nuclear interactions of masses, considering it, as well as the Hubble radius $R_H$ of the Universe, as a universal parameter together with $\hbar$ and $c$. We believe that the introduction of the four basic fundamental parameters $\hbar$, $c$, $m_Q$ and $R_H$ suffices for the representation of all basic interactions in the Universe at the phenomenological level.

In accordance with the Dirac's idea, the introduction of the value $R_H$ or the Hubble constant $H$, along with $\hbar$, $c$ и $m_Q$, means dependence these basic fundamental constants on the global time. Each $H$ value corresponds to a specific epoch in the evolution of the Universe and the well-defined set of fundamental constants $\hbar$, $c$ и $m_Q$ in this era.

In this case, the potential energy of interaction between the two masses $m_1$ and $m_2$ at the distance $\rho$ apart ("the Law of Universal Gravity") can be conveniently represented in the form:

$$V_g(\rho) = -G\frac{m_1 m_2}{\rho} = -\frac{q_g^2}{\rho} \mu_1 \mu_2, \quad \mu_i \equiv \frac{m_i}{m_Q}. \qquad (5a)$$

The quantity $q_g^2 \equiv G m_Q^2$ is considered as the squared "elementary gravitational charge". We introduce also a dimensionless constant $\alpha_g$ of the gravitational interaction, defined as:

$$\alpha_g = \frac{G m_Q^2}{\hbar c} = (2\pi)^2 \frac{\hbar H}{m_Q c^2} = 2^{3/2} \pi^2 \frac{a_Q}{R_H} \approx 2.85 \cdot 10^{-40}. \qquad (6)$$

The expression (6), which reflects the quantum essence of gravitational interaction ($\alpha_g \sim \hbar$) allows understanding the nature of the unique smallness of gravitational interaction, reflected in the value of the dimensionless constant $\alpha_g$, which is by 38 orders of magnitude smaller than the fine-structure constant $\alpha_e$. From the expression (6) it becomes clear that the formal reason for this difference is the smallness of the ratio of the characteristic size $a_Q$, corresponding to the mass $m_Q$, a kind of "seed" of the gravitational interaction, to the characteristic size of the Universe. Thus, the name of "the Law of Universal Gravity" for the relation describing the gravitational interaction of two arbitrary masses is "justified" at the conceptual level.

Let us mention, for comparison, the value of the squared "elementary charge of the weak nuclear interaction" $q_F^2 \equiv G_F/a_Z^2$, where $a_Z = 2^{1/2}\hbar/m_Z c \approx 3.3 \ 10^{-16}$ cm is the Bohr radius, connected with the mass of the intermediate $Z^0$ vector boson, $m_Z = 91.2$ GeV/s$^2$ = $1.62 \ 10^{-22}$ g, and $G_F = 1.436 \cdot 10^{-49}$ erg·cm$^3$ is the Fermi constant [16] of the four-fermion interaction. In this case, we get for the value of the corresponding dimensionless constant of the weak nuclear interaction:

$$\alpha_F = \frac{q_F^2}{\hbar c} = \frac{a_F^2}{a_Z^2} \approx 5.7 \ 10^{-2}, \tag{7}$$

where $a_F = (G_F/\hbar c)^{1/2} \approx 0.7 \cdot 10^{-16}$ cm. Note that $\alpha_F = \alpha_e = 1/137 \approx 0.73 \cdot 10^{-2}$, if we choose $m_W^* = 35.2$ GeV/s$^2$ as a basic mass of the electroweak interaction.

The usage of the introduced collection of the basic universal constants $\hbar$, $c$, $m_Q$ and $R_H$ allows presenting in a more "compact" fashion not only the gravitational constant $G$ and the dimensionless constants of the gravitational and weak nuclear interaction, but also the Planck numbers (1) and (2):

$$a_{Pl} = 2\pi a_Q \left(\frac{a_Q}{R_H}\right)^{1/2}, \quad t_{Pl} = 2\pi \tau_Q \left(\frac{a_Q}{R_H}\right)^{1/2}, \quad m_{Pl} = \frac{1}{2\pi} m_Q \left(\frac{R_H}{a_Q}\right)^{1/2}, \quad w_{Pl} = \frac{m_Q c^2}{4\pi^2 \tau_Q} \cdot \frac{R_H}{a_Q}, \tag{1a}$$

where $\tau_Q = a_Q/c \approx 0.43 \ 10^{-23}$ s is the time scale, corresponding to the space scale $a_Q$. The relations (1a) clarify the cosmological essence of the "smallness" of the Planck parameters $l_{Pl}$ and $t_{Pl}$, as well as the cosmological scale of the quantities $m_{Pl}$ and $w_{Pl}$, thus demonstrating the heuristic justification of the representation (4) for the Planck constant. Therefore, in spite of the transcendent character of the relations (1a) and (6) − the principal impossibility of obtaining direct experimental information on the character of the enclosed in them dependencies on the ratio $R_H/a_Q$ – the indicated relations, in contrast to (1), (2) and (4), should be viewed not as "arithmetical" relationship but as the phenomenological relationship, which in the future may become not only the basic reference points, but the basic postulates for the deductive construction of theoretical models of the dynamics of the Universe.

Despite the above, the developed phenomenological approach allows us to move forward also in the clarification of the physical reasons for the indicated smallness of the dimensionless constant $\alpha_g$ of the gravitational interaction: $\alpha_g \sim a_Q/R_H$. To do this, one should introduce one more phenomenological image: an idea about the electromagnetic component of the physical vacuum, "EM vacuum", as the basic medium of Universe and the "Casimir polarization" of this medium in the vicinity of the atomic nuclei of all material objects in Universe (see also [3, 13]).

3. ELECTROMAGNETIC COMPONENT OF THE PHYSICAL VACUUM, EM VACUUM, AS THE BASIC MEDIUM FOR GRAVITATION

It is known [17] that EM vacuum manifests itself through the fluctuating mean values of the squared intensity of the electric and magnetic fields. This means that a noise electrodynamic component acts from EM vacuum upon the system of concentrated charges and local currents of any material body [18]. The power spectrum of this component, as it should be assumed, corresponds to the "white noise". Such stochastic effects are known (see, for example, [19]) to initiate smearing-out of the point electrons, which leads to the appearance of the natural width of the excited levels of atoms, determines the Lamb shift and a number of other phenomena.

The electromagnetic vacuum, as well as electromagnetic radiation, fully manifest themselves also in the electronic and nuclear subsystems. It is exactly the interaction of the electronic subsystem of the excited atom with the zero-point oscillations of the electromagnetic field in vacuum, the frequency of which is equal to the frequency of the emitted quantum, that is the initiating factor, the cause of its spontaneous emission (the states of the excited atom cease to be stationary).

The initiating factor of the zero-point oscillations of the electromagnetic field is also manifested in the spontaneous emission of γ-quanta by the excited nuclei. Thus, by analogy with the induced emission of excited atoms, the extraneous impact of an outburst of X-ray radiation on a set of excited iron-57 nuclei trapped in the structure, which is an X-ray waveguide (the atoms are at the crest of the passing excitation), leads to a six-fold decrease in the time of emission of γ-quanta by the iron-57 nuclei [20].

It is obvious that in the interaction of EM vacuum with a system of charges and local currents of each material body, at the interphase boundary between EM vacuum and the surface of the body, there are formed certain boundary conditions for the tangential and normal components of the electric field strength, on which the respective impacts of EM vacuum on the bodies depend. In particular, in the vacuum gap of the width $d$ between the perfectly smooth metal plates (assuming that the tangential component of the strength vector $E_\tau$ of the electric field, created by the zero-point oscillations of EM vacuum on the surface of the plates is equal to zero), there arises a force of attraction (the "Casimir force"), the value of which is determined by the increase in the gap $d$ of one "resonant" frequency of EM vacuum, equal to $c/d$, and the suppression of a wide range of other frequencies. This force per unit area of the plates (pressure) is given by [21, 22]:

$$F_C(d) = -\frac{\pi^2}{240} \cdot \frac{\hbar c}{d^4}. \tag{8}$$

The validity of the relation (8) is established experimentally for the variation of the gap width $d$ from 50 to 500 nm [21].

Guided by the relationship (8), Casimir expressed the idea (see [23]) that the corresponding forces should stabilize elementary particles as well, in particular, electron, keeping a negative charge in it. In a number of subsequent model calculations of such Casimir forces, a particle of the mass $m_0$ was represented by a sphere of radius $a$ with metallic conductivity, so that the tangential component $E_\tau$ of the strength vector of the electric field created by the zero-point oscillations of the EM-vacuum on the surface of the particle, assumed to be zero, $E_\tau = 0$ (the boundary condition of the second kind). However, the relevant calculations made using different methods of renormalization have shown [23, 24] that the vacuum energy $U_a$ of such a sphere of radius $a$, defined up to a certain constant ($\overline{E}_0$), formally infinite and independent of the sphere radius, is positive and equal to $U_a = 0.04618\,\hbar c/a$. Note, the similar result could be obtained by using the method of calculation of the Casimir force without reference to the EM vacuum and the procedure of renormalization [22]. In accordance with [22], the vacuum-to vacuum graphs with the inner lines, that define the zero-point energy, did not enter the calculation of the Casimir force, which instead only involves graphs with external lines.

The positive value of the calculated energy means that, inside such a sphere, there acts the repulsive forces and the sphere seeks to expand. Therefore, the Casimir's idea as applied to solving the problem of particle stability was rejected. However, this idea has received a new development on the basis of the phenomenological concepts developed in [2, 13, 25]. To be specific, we will consider as a material object the atomic nucleus and *a priori* suppose that the energy of this object determined by the Casimir forces is negative. Casimir force arises from the action of energy-containing EM vacuum on an atomic nucleus treated as a set of distributed charges (quarks in nucleons) and local exchange currents. These effects are manifested in the boundary condition as the coupling of components of the EM vacuum electric field vector on the outer and inner boundaries of the atomic nucleus (see below). In fact, we believe that EM vacuum in the vicinity of the material object is polarized: the oscillation spectrum of EM vacuum is rearranged, and a unified dynamic system is formed composed by the atomic nucleus and vacuum, an "EM vacuum polaron". The introduction of the EM vacuum polaron allows qualitatively understanding the genesis of *a priori* limitation of the speed of moving material objects with nonzero rest mass by the velocity of light $c$ in EM vacuum which is equal to $c = 3 \cdot 10^{10}$ cm/s. This conclusion is due to the limitation of the rate of reorganization of EM vacuum adjacent to all atomic nuclei of the moving object. It is clear that the rate of such reorganization of the physical vacuum areas adjacent to the boundaries of these nuclei is limited by the speed of light $c$, which makes impossible the movements of material bodies with such speeds in the EM vacuum, which acts as a material medium, a kind of "ether".

It is obvious that if we present the atomic nucleus as a vacuum polaron, then the vacuum energy $U_a$ of the sphere, the boundary of the nucleus, is negative, and the normal component of EM vacuum pressure outside the nucleus exceeds the corresponding value inside the nucleus. In this case, the boundary condition in the calculation of the value of $U_a$ should be different: it should be a condition of the third kind, such as $E_n = \chi \, dE_n/dr$, where $\chi$ is a phenomenological parameter depending on the state of nuclear matter and characterizing the dynamics of the processes of "matching" the normal components $E_n$ on both sides of the boundary. We will consider $\chi$ as a parameter that defines the stationary state of the nucleus with the dynamic unity of the system "nuclear matter – the electron subsystem of the atom". For certain values of this parameter, $\chi > \chi_0$, the vacuum energy of such spherical object is negative, $U_a < 0$, whereas the object itself is stable. We believe that it is exactly due to such dynamic relationship of the internal region of the atomic nucleus with the external basic medium that there is formed an internal stable organization of nuclear matter as a set of interacting nucleons. Since for $\chi = 0$, in accordance with the above results [23, 24], the vacuum energy value $U_a > 0$, then for some value of the parameter $\chi_0 > 0$ there is achieved $U_a = 0$, and $U_a > 0$ for $\chi < \chi_0$. Therefore, for the nucleus to lose stability, it suffices to introduce into nuclear matter an excitation, ensuring the equality $\chi = \chi_0$ [25].

Let us indicate also that another justification of the introduction of the vacuum polaron hypothesis can be the possibility of the understanding, on this basis, the quantum essence of the "wave-particle" image [13]. Indeed, the movement of the particles with the mass $m$ in EM vacuum with the velocity $u$, in fact, means the displacement of the local heterogeneity of EM vacuum with the momentum $p = mu$ and the energy $E = \sqrt{p^2 c^2 + m_0^2 c^4} = mc^2 = \hbar \omega$, where $m_0$ is the mass of the particle for $u = 0$ and $\omega$ is the cyclic frequency of the moving perturbation. The phase velocity $u_{ph}$ of the displacement of such perturbation of EM vacuum with the wave number $k = p/\hbar = 2\pi/\lambda_{dB}$, which we connect with the de Broglie wave ($\lambda_{dB}$ is the corresponding wavelength), is equal to $u_{ph} = \omega/k = E/p = c^2/u$, whereas the group velocity of the displacement equals $u_g = dE/dp = u$. In this case, the domain of the EM vacuum Casimir polarization is "attached" to the moving particle, and the phase velocity $u_{ph}$ of the moving heterogeneity region is uniquely related to the particle velocity $u$. Therefore, the spreading of the domain of the moving perturbation of the EM vacuum will not occur when the perturbation propagates (there is no time dispersion).

These wave properties are inherent in each of the moving particle. It is by this fact that the known phenomena are conditioned of the formation of the diffraction patterns for extremely low fluxes of particles, striking on the diffraction grating, when the interaction between the particles is impossible. It also becomes clear the nature of the impossibility of simultaneous determining the position of a particle in space and its momentum. Indeed, the exact value of the position of the center of mass of the particle can be determined only for a particle at rest relative to EM vacuum, because when the particle moves, not only the region of Casimir polarization of EM vacuum in the particle neighborhood is deformed, introducing uncertainty into the position of its center of mass, but also the very mass of the particle changes (see the discussion of the last issue in [2, 3, 13]).

When considering EM vacuum as the basic medium for the movement of all material bodies in the entire space of the Universe, it is natural to associate exactly with EM vacuum the base inertial system, common for all masses of the expanding Universe. It is also natural, following Burlankov [26], to adopt for this system a global time scale $t$, common for all points of the Universe and counted from the time $t = 0$, corresponding to the Big Bang. It is to this, physically distinguished system that the equations of the dynamics of the Universe are "tied" (the Friedmann equations [27-29]), in which the entire mass and energy, as well as the associated with this system EM vacuum are assumed to be uniformly distributed throughout the volume of the Universe. We define this frame of reference as a "system of Mach", who first introduced a basic inertial system in which "a material particle does not move in unaccelerated motion relatively to space, but relatively to the center of all the other masses in the Universe" ([6], p. 60).

In connection with the binding of the studied objects to the inertial and non-inertial reference systems, it should be noted that this practice has proven its practicality since the times of G. Berkeley ([30]), who proposed to connect a choice of inertial reference systems with the orientation of the coordinate axes to the "motionless stars". The current practice shows that it is possible to introduce virtually absolute inertial frame of reference in which the relative acceleration of the sufficiently distant from each other bodies is no greater than $10^{-10}$ m/s² [31]. Such a system, in particular, is the International Celestial Reference System, for which the origin is taken to be the position of the barycenter, the center of mass of the solar system, whereas its axes are fixed in space with respect to quasars, the most distant objects of the observable Universe. Thus, the dynamics of sufficiently large volumes of the Universe can be considered in the long-length frames, which with good reason can be considered inertial.

In concluding this section, we note that the introduction of the EM vacuum as an active basic medium, which is polarized in the vicinity of material objects, allows approaching the understanding of the physical nature of the basic postulates of quantum mechanics. We should note this referring to the remark by R. Feynman: "I can safely say that nobody understands quantum mechanics" ([3], p. 129). On the other hand, the introduction of a basic inertial reference system associated with the EM vacuum resolves some of the issues that arise [32] in connection with adoption of the special relativity postulates (see also [3] and section 4 of this article).

## 4. EM VACUUM IN THE FORMATION OF THE GRAVITATIONAL ATTRACTION OF MASSES

According to [3, 13], it is exactly to the Casimir polarization of EM vacuum in the vicinity of any material objects in the Universe that there should be connected the phenomenon of gravity: the attraction of these objects in accordance with the Law of Universal Gravitation. The expression for the corresponding potential energy of the Casimir field in the neighborhood of a material object (for definiteness, a particle with the mass $m_0$) has the form:

$$U(\vec{r}) = -\gamma_0 \frac{\hbar c}{r}. \qquad (9)$$

Here $\vec{r}$ is the radius-vector (we connect the coordinate system with EM vacuum and assume that the particle at rest is localized at the origin); $\gamma_0$ is a dimensionless parameter, characterizing the intensity of the introduced interaction.

The solution of the Schrödinger equation in a centrally symmetric field with potential energy in this form is well known [33]. The energy levels $\overline{E}(n_r)$ of the discrete spectrum, reflecting the degree of interconnection of the particle of the mass $m_i$ with EM vacuum due to its polarization, and the corresponding expression $a_B$ for the radius $a_{Vi}$ of the domain of the EM vacuum Casimir polarization in the neighborhood of the particle $i$, have the form:

$$\overline{E}(n_r) = -\gamma_0^2 \frac{m_i c^2}{2 n_r^2}, \qquad (10)$$

$$a_{Vi} = \frac{2\hbar}{\gamma_0 m_i c}. \qquad (11)$$

Here $n_r$ is the principal quantum number. For $\gamma_0 = \sqrt{2}$, the position of the lower energy level ($n_r = 1$), characterizing the binding energy of the considered particle with EM vacuum, corresponds in the absolute value to the "rest energy of the considered particle" $\overline{E}_0 = m_0 c^2$ in the form suggested by Einstein, where $U(\vec{r})|_{r=a_B} = -m_0 c^2$. Within the developed phenomenological ideas, the quantity $\overline{E}_0$ is defined as the "binding energy of the particle with EM vacuum", so that the mass defect in the nuclear processes just characterizes the energy released due to the difference of binding energies of the initial and final products with EM vacuum.

According to (11), the value of the domain of the EM vacuum Casimir polarization in the neighborhood of the proton equals $a_{Vp} = 2.82 \cdot 10^{-14}$ cm, i.e. corresponds to the action scale of the nuclear forces. Thus, the quantity $q_s^2 \equiv \gamma_0 \hbar c$ for $\gamma_0 = \sqrt{2}$ in the expression (9) can be defined as the squared "elementary charge of strong interaction", so that the dimensionless constant $\alpha_s$ of such interaction, by analogy with the fine structure constant $\alpha_e$ and the introduced above constants $\alpha_g$ and $\alpha_F$, can be naturally represented in the form

$$\alpha_s = q_s^2 / \hbar c = \sqrt{2}. \qquad (12)$$

The value of the fine structure constant $\alpha_e$ can be written as

$$\alpha_e = \frac{a_{Ve}}{a_B} \alpha_e = \frac{a_{Ve}}{\sqrt{2} a_B} = \frac{1}{137}. \qquad (12a)$$

Here, $a_{Ve} = 5.2 \cdot 10^{-11}$ cm is the radius of the domain of EM vacuum Casimir polarization in the neighborhood of the electron; $a_B = \hbar^2 / m_e e^2 = 0.52 \cdot 10^{-8}$ cm is the characteristic size of the simplest hydrogen atom, its "Bohr radius". The relation (12a) essentially characterizes the measure of overlapping of the Casimir polarization domains of the nucleus and the electronic subsystem of the atom.

If a particle possesses a structure (hadron), then the dependence (9a) inside such particle can be treated as a "seed" potential energy of the nuclear forces, which are characterized by the "nuclear charge" $q_s$ and the strong coupling constant $\alpha_s$. It is obvious that, as a result of dynamic mobility of the nuclear matter "inside" such particle, the seed potential of the nuclear forces is shielded, and the effective potentials are formed of the "short-range" nuclear forces, exponentially declining with the distance, of the type of the Yukawa potential

$$U(\vec{r}) = -\frac{q_s^2}{r} \exp(-a_{Bs} r) \qquad \text{for } r < a_B, \qquad (9a)$$

which should be introduced instead of (9). This corresponds to the standard idea about the dynamic nature of the nuclear forces, which are usually related to the $\pi$-meson exchanges by nucleons.

The unexpectedness of the last result is in the effectively revealed at the phenomenological level physical unity of the electromagnetic and strong interactions. Since the nature of the Casimir effect is related to the local spatial charges of the electromagnetic component of the physical vacuum in the vicinity of and inside a material object (with the resonant amplification of the frequencies, characteristic for the structure of the object, and the exclusion of other frequencies, contained in the spectrum of the physical vacuum), then the nuclear forces in such a phenomenological model are a response of the nuclear matter on the spatial-temporal nuclear scale to the action of the electromagnetic component of the physical vacuum. Obviously, such a response is extremely specific and depends both on the specific character of configuration and dynamics of each individual nucleus and its excitation. From this point of view, one can qualitatively understand the usually discussed dependence of the strong coupling constant $\alpha_s$ on the excitation value [15, 16].

The obtained result (10) can be regarded as a kind of "justification" of the use of (9) up to the distances corresponding to the sizes of elementary particles, since it leads to physically meaningful results. First of all, it turns out that all objects of our world are connected with the EM vacuum, considered as a physically distinguished basic environment. It is also becomes clear the physical cause of the appearance in the expression for the "particle rest energy" $E_0$ of a characteristic of this basic environment, the speed of light in vacuum as a parameter, included in the definition of the potential energy (9), characterizing, in the product with the Planck constant, the Casimir polarization of EM vacuum and the corresponding reorganization rate of EM vacuum in the vicinity of the particle as it moves in this basic medium. It is this polarization of the particle in EM vacuum acting as a factor of stabilization of elementary particles and stable isotopes, that may keep from breakup the electron, as well as others elementary particles having a charge [34]. It is possible that the nature of quark confinement inside hadrons [16] should also be linked to the Casimir force. Of interest also are the excited states of the particle localized in the vacuum. In particular, these levels can be manifested as "resonances", the short-lived excited states of hadrons with the characteristic lifetimes of $10^{-22} - 10^{-24}$ s, which are formed in the interaction of π-mesons with nucleons [35].

We should note the same character of the dependence of the Casimir potential energy (9), characterizing the polarization of EM vacuum in the vicinity of particles with different masses $m_i$, on the value of the radius-vector $r$. At the same time, the "depths of occurrence" of the lower energy levels in such potential wells, characterizing the binding energy of different masses $m_i$, are different. In this case, in each well there may localize a depending on $m_i$ finite number of "excited" levels ($n_r > 1$), if we have in mind their number from the basic state to the level of external "noise" (for example, the heat one, which is equal to $k_B T$). Taking the radius $a_{Vi}$ (11) as the boundary with EM vacuum of the considered material object, we identify on the both sides of this boundary the "external" and "internal" regions of the introduced vacuum polaron. The EM quanta at the excited levels, localized in the "external" region of polaron, should be referred to as virtual quanta, for which the wave vector $\vec{k}$ and frequency $\omega$ are independent variables, not connected by the dispersion relation $\omega = kc = 2\pi c/\lambda$, which holds for the real photon. To each excited level, there correspond the virtual EM quanta with the frequencies $\overline{\omega}(n_r) = \overline{E}(n_r)/\hbar$. Moreover, there may "condensed" at each level an arbitrary number of virtual EM quanta.

Due to the infinite range of action of potential energy (13), the presence in the medium of various material objects with the masses $m_i$ inevitably leads to overlapping of the potential fields and forming the fields of the particle attraction – gravity fields. The gravitational energy of the system $N$ of attracting masses, which are distributed in the EM vacuum, in general, un-uniform, can be characterized by the local value $U(\vec{\xi}; \vec{r}_1, \vec{r}_2, ..., \vec{r}_N)$ at the radius vector $\vec{\xi}$, where $\vec{r}_i$ ($i = 1, 2, ..., N$) is the radius vector of a particle with the mass $m_i$. Consider for example the corresponding Casimir potential energy $U(\vec{\xi}; \vec{r}_1, \vec{r}_2)$ of two particles with the masses $m_1$ and $m_2$. As before, we associate the coordinate system with EM vacuum. Then we have:

$$U(\vec{\xi}; \vec{r}_1, \vec{r}_2) = -\frac{\sqrt{2}\hbar c}{|\vec{\xi} - \vec{r}_1|} - \frac{\sqrt{2}\hbar c}{|\vec{\xi} - \vec{r}_2|}. \tag{13}$$

Introduce the radius-vector $\vec{R}$ of the center of mass of the particles $m_1$ and $m_2$, as well as the radius-vector $\vec{\rho}$ of the difference of the radius-vectors $\vec{r}_1$ and $\vec{r}_2$:

$$\vec{R} = m_{12}\left(\frac{\vec{r}_1}{m_2} + \frac{\vec{r}_2}{m_1}\right), \quad \vec{\rho} = \vec{r}_1 - \vec{r}_2, \quad m_{12} = \frac{m_1 m_2}{m_1 + m_2}.$$

Then

$$U(\vec{\xi}; \vec{r}_1, \vec{r}_2) = -\sqrt{2}\hbar c \left[\left|(\vec{R}-\vec{\xi}) + \frac{m_{12}}{m_1}\vec{\rho}\right|^{-1} + \left|(\vec{R}-\vec{\xi}) - \frac{m_{12}}{m_2}\vec{\rho}\right|^{-1}\right] = $$
$$= -\sqrt{2}\frac{\hbar c}{m_{12}}\left[m_1\left|\vec{\rho} + \frac{m_1}{m_{12}}(\vec{R}-\vec{\xi})\right|^{-1} + m_2\left|\vec{\rho} - \frac{m_2}{m_{12}}(\vec{R}-\vec{\xi})\right|^{-1}\right]$$
(13a)

In the framework of the developed phenomenology, gravity is the manifestation of interaction between each pair of material bodies as a result of Casimir polarization of the EM vacuum in their neighborhood. Therefore, when considering the dynamics of a system of bodies, determined by the collection of all paired gravitational interactions, one must take into account the changes of the positon of the center of mass of each pair of interacting bodies relative to the fixed center of mass of the entire system. If we are only interested in the potential energy of attractive interaction between two particles, then for the exclusion from consideration of the dynamics of the system as a whole there must be chosen $\vec{\xi} = \vec{R}$. We obtain in this case:

$$U(\vec{r}_1, \vec{r}_2)\big|_{\vec{R}=0} = -\frac{\sqrt{2}m_1 m_2}{m_{12}^2} \cdot \frac{\hbar c}{\rho}.$$
(13b)

We should keep in mind that the Casimir form of the potential energy (9) of the EM vacuum polarization by a particle of the mass $m_i$ is valid only in the immediate vicinity of the particle. Therefore, at the macroscopic distances, for $r \gg a_B$, there must be manifested for each of the considered particles with the masses $m_1$ and $m_2$ the influences on these masses not only from the relatively proximate particles, but also from the distant masses, and, in accordance with a Mach's idea [36], from all masses of the Universe. These effects must also appear in the expression for the potential energy (13b). Since the product $(m_1/m_{12}) \times (m_2/m_{12})$ is included in this expression, and our concern is the potential energy of interaction between two particles with the masses $m_1$ and $m_2$, then we will keep the product $m_1 \times m_2$ in the numerator of (13b). At the same time, the averaged impact of all the masses of the Universe on the potential energy of interaction between two distinguished masses $m_1$ and $m_2$ will be taken into account by replacing the squared mass $m_{12}^2$ in the denominator of (13b) by the squared mass $m_x^2$ and averaging (13b) over all possible values of the mass $m_x$, from a certain minimum mass $m_{min}$ to infinity.

Introduce a distribution function $f(m_x)$ for the contribution of the masses of extraneous objects to the potential energy of interaction of the masses $m_1$ and $m_2$ and choose, following [37], the simplest approximation for $f(m_x)$, characterized by self-similarity which is specific for the cosmological objects:

$$f(m_x) = (q-1)\frac{m_{min}^{q-1}}{m_x^q}; \quad \int_{m_{min}}^{\infty} f(m_x)\, dm_x = 1.$$
(14)

Here $q$ is a distribution parameter (for the estimates, in view of [37], we put $q \approx 2$). Then, after the indicated averaging of the expression (13b) for the potential energy $V_{12}(\rho)$ of gravitational attraction of two masses $m_1$ and $m_2$ at the distance $\rho$ from each other, we get:

$$\overline{U}(\vec{r}_1, \vec{r}_2)\big|_{\vec{R}=0} \equiv V_g(\rho) = -\sqrt{2}m_1 m_2 \frac{\hbar c}{\rho}\int_{m_{min}}^{\infty}\frac{f(m_x)\,dm_x}{m_x^2} = -\sqrt{2}\frac{q+1}{q-1}\cdot\frac{m_1 m_2}{m_{min}^2}\frac{\hbar c}{\rho} = -\sqrt{2}\frac{m_1 m_2}{m_M^2}\frac{\hbar c}{\rho}.$$
(15)

Here we introduce the mass

$$m_M \equiv \sqrt{\frac{q-1}{q+1}}\, m_{min},$$
(16)

which is defined as the Mach mass. Comparing the dependence (15) with Newton's law (5) and taking into consideration (1), we find:

$$m_M = \sqrt{2}\, m_{Pl}, \qquad (16a)$$

whereas, in view of (1a), we get for the gravitation constant $G$:

$$G = \sqrt{2}\,\frac{\hbar c}{m_M^2} = \frac{\hbar c}{\sqrt{2}\, m_{Pl}^2} = 2^{3/2}\pi^2\,\frac{\hbar c}{m_Q^2}\,\frac{a_Q}{R_H}, \qquad (17)$$

in accordance with the obtained earlier phenomenological expression (6) for the dimensionless constant $\alpha_g$ of the gravitational interaction. Obviously, in case of a system of gravitationally interacting masses, the potential energy of each of the pairs of masses $m_1$ and $m_2$ should be represented, taking into account (13a), (13b), (16a) and (17), in the following form:

$$U(\vec{\xi};\vec{r}_1,\vec{r}_2) = -\sqrt{2}\,\hbar c\,\frac{m_{12}}{m_M^2}\left[ m_1\left|\vec{\rho} + \frac{m_1}{m_{12}}(\vec{R}-\vec{\xi})\right|^{-1} + m_2\left|\vec{\rho} - \frac{m_2}{m_{12}}(\vec{R}-\vec{\xi})\right|^{-1}\right] =$$

$$= -G m_{12}\left[\frac{m_1}{\left|\vec{\rho} + \frac{m_1}{m_{12}}(\vec{R}-\vec{\xi})\right|} + \frac{m_2}{\left|\vec{\rho} - \frac{m_2}{m_{12}}(\vec{R}-\vec{\xi})\right|}\right], \quad m_M = 2^{1/4}\left(\frac{\hbar c}{G}\right)^{1/2}. \qquad (13c)$$

The expression (17) elucidates the physical essence of the gravitational constant $G$ and its relatively small value due to the small spatial extent of the region of Casimir polarization of EM vacuum in the vicinity of gravitating masses (the parameter $a_Q$) as a source of gravitational interactions and its cosmological character (the parameter $R_H$). In fact, the made conclusion about a single normalization mass $m_M$ in the expression for the potential energy of attraction of any two masses $m_1$ and $m_2$ can be regarded as a representation of the Mach's idea [36] on the impact of all the physical bodies of the Universe on each specific mass. Indeed, because of the fundamental cosmological principle – homogeneity and isotropy of the mass distribution in the Universe − the resulting component of the force of attractive influence of all the masses of the Universe on each of the masses $m_1$ and $m_2$ in any distinguished direction, including the one connecting these masses, contains both an amplifying factor (in the denominator of expression (17) there is a square of a small mass, $m_M \sim 10^{-5}$ g, serving effectively as a reduced mass for the interacting macroscopic masses), and a factor reflecting the cosmological scales in the distribution of the attracting masses, the factor which is very small but still non-zero due to finiteness of the ratio $a_Q/R_H$. It should be noted now that in this case we are talking about the impact on each mass from all the masses of the Universe in connection with the identification of the genesis of the law of interaction of gravitating (gravitational) masses, not about the role of these interactions in the manifestation of the inertia properties at the non-accelerated motion of a point particle relative to the center of mass of all the masses of the Universe, as it seemed to Mach. With regard to the manifestation of inertia by the gravitating mass, according to [13], there is revealed in the genesis of this phenomenon, first of all, the direct role of the EM vacuum.

It should be pointed out that in [12] a somewhat different representation for the gravitational constant $G$ in the Law of Universal gravitation was proposed (5a). If we use the expression (12) for the dimensionless constant $\alpha_s$ of strong interaction, then the expression (5a) for Newton's law can be written as:

$$V_g(\rho) = -\frac{q_g^2}{\rho}\mu_1\mu_2 = -\frac{\alpha_s}{\eta_g \rho}\mu_1\mu_2. \qquad (5b)$$

Here we introduced the quantity:

$$\eta_g = \frac{m_M^2}{m_Q^2} = \frac{2 m_{Pl}^2}{m_Q^2} \approx 0.46\cdot 10^{40}, \qquad (18)$$

which can be defined as a "gravitational permeability of the EM vacuum". The anomalously large value of the introduced parameter $\eta_g$ may mean that, in accordance with the idea of Mach, the masses of the Universe that are "cosmologically distant" make a contribution to the "gravitational interaction" of the two considered masses.

The representation of Newton's law using the dimensionless constant $\alpha_s$ of the strong interaction (12) and the masses $m_1$ and $m_2$, normalized by the mass $m_Q$ which is characteristic for the nuclear matter, points out more explicitly to the "intranuclear" processes, which determine through the appropriate boundary conditions the Casimir polarization of EM vacuum "outside the nucleus", which is the source of the gravitational attraction of masses. This actually means that gravity is the manifestation of the strong interaction at large distances significantly exceeding the characteristic nuclear scales. .

## 5. EM VACUUM IN THE FORMATION OF THE RELATIVISTIC EFFECTS AND THE PHENOMENON OF INERTIA

Introduction of ideas on the Casimir polarization of the EM vacuum in the vicinity of an atomic nucleus as a "vacuum polaron" allows to understand the nature of inertia, as well as known relativistic effects – the increase in the mass of a particle with an increase in its velocity and an increase in the lifetime of decaying relativistic particles. We will assume, following the classics of physics of the XIX century, that the relativistic effects under consideration, primarily the dependence of the particle mass on velocity, are determined by local reconstructions of the base medium in the vicinity of particles whose velocity is comparable with the speed of light $c$ in vacuum. If the "ether" was postulated as the basic medium in the science of the XIX century, then according to the arguments presented in this paper, the EM vacuum should be considered as the base medium of the Universe, having in mind the possibility of local rearrangements of the domain of the EM vacuum Casimir polarization in the vicinity of a material particle at its velocities close to the value of $c$. It is then possible to consider relativistic displacements of macroscopic objects, since the domains of the EM vacuum Casimir polarization in the vicinity of the atomic nucleus and the electron subsystem of every atom do not overlap. Indeed, as indicated above, these domains are of the order of $10^{-13}$ cm and $10^{-11}$ cm, respectively, whereas the sizes of atoms are of the order of $10^{-8}$ cm. For this reason, the postulated changes in the geometry of the domains of the EM vacuum Casimir polarization of the nuclei and the corresponding electronic subsystems of all the atoms of an arbitrary object, including any macroscopic body, at its relativistic velocities occur independently.

Let us consider these changes in more detail by the example of the relativistic change in the domains of the Casimir polarization in the vicinity of the atomic nucleus. Let a nucleus move freely with the velocity $u$ relative to the introduced basic inertial reference system, and let it have the mass $m_0$ for $u = 0$. Let us find out with what changes in the polarization of the domain of the EM vacuum in the vicinity of this material particle there can be linked the appearance of the factor $\eta_u$ in the expression for the total energy $E_u$ of the particle [38]:

$$E_u = \eta_u m_0 c^2 , \quad \eta_u = \left(1 - \frac{u^2}{c^2}\right)^{-1/2} , \qquad (19)$$

For the first time this factor was introduced by Oliver Heaviside in 1889 (we quote [39], p. 35) in the model calculations of the aether drag (a basic medium of the science of the XIX century) by the moving charged spherical particle of the radius $a$ and the mass $m_0$. The effect of the aether drag increased proportionally to $\eta_u$ with increasing of the particle velocity due to the geometric displacement of the "Faraday tubes" toward the equatorial plane passing through the center of the sphere and perpendicular to the direction of movement. In this case, the Faraday tubes were initially oriented normally to the surface of the particle. If we conditionally restrict the domain of such system by a sphere of radius $R$, equal to several radii $a$, then, under a small velocity of the particle, the initial spherical geometry of the system is transformed into a flattened ellipsoid of revolution around the minor axis of the ellipse oriented along the particle path, so that

$$R_s = R \cdot \eta_u^{-1} < R_l = R. \tag{20}$$

Here $R_s$ and $R_l$ are the sizes of the small and major semi-axes of the oblate ellipsoid of rotation, respectively. Based on this result, J.J. Thomson [39] calculated the kinetic momentum in the space surrounding this particle and showed that the mass of the particle, with its velocity increasing, grows proportionally to the factor $\eta_u$ due to increasing of the total amount of aether entrained by the Faraday tubes in the movement of the charged particle. The subsequent formation of the special theory of relativity (STR) and the experimental studies have fully confirmed the universal role of the factor $\eta_u$ in the entire the spectrum of relativistic phenomena.

In accordance with the concepts being developed, the EM vacuum, being the basic medium and a physically distinguished system for all the objects in our Universe, is a modern analogue of the aether of the science of the XIX century. Thus, it is natural to try to relate the effect of the mass increasing of a relativistic particle in the STR with the changes in the vacuum polarization region in the vicinity of such particle in the direction of its motion, following the general ideas of O. Heaviside and J.J. Thomson. It should be understood that the postulated vacuum polarization in the vicinity of any material object with the EM vacuum, in reality, implies openness in the dynamic sense of every atomic nucleus of this object for EM vacuum in accordance with the boundary conditions introduced at the interface of every nucleus with EM vacuum. In other words, the properties of any elementary particle are formed in the interaction of its inner essence with the electromagnetic component of the physical vacuum [25].

We will assume that during non-relativistic movement of the considering atomic nucleus, relative to the basic inertial reference system, there is realized "equilibrium" exchange of virtual photons, which are localized in the EM vacuum polarization domain in the vicinity of the nucleus, and of the virtual photons of the EM vacuum as the basic medium. Furthermore, the polarization region is characterized by a certain level of "solvation", the condensation of virtual photons on the system of excitation levels of the nucleus. The exchange of virtual photons, realized on the boundary "atomic nucleus – EM vacuum" is disturbed under external influences on the nucleus (see Section 6) or under its relativistic movements.

In accordance with the logic of Weizsacker [40], the very fact of any change in the state of a system, in this case, due to the transition of virtual photons, is inevitably associated with the dissipativity and irreversibility of such process. But to what extent in the influence of quantum fluctuations of the EM vacuum − the fluctuating mean square values of the intensity of electric and magnetic fields [17] − there can manifest the real energy to initiate irreversible dissipative processes? In the author's opinion, as a basis for the possibility of this type of processes one can consider the established role of quantum fluctuations of the EM vacuum in the formation of the radiation pressure (the macroscopic manifestations of these effects were observed in [41]), as well as the static Casimir effect [21] together with the dynamic Casimir effect [42, 43] with the direct conversion of the fluctuations of virtual photons into real photons in the region of the boundaries of the objects moving with relativistic velocities.

To characterize the considered exchange of virtual photons, we consider, in accordance with [44, 45], a boundary condition of the third kind with the introduction of a boundary ad-state or a state I, from which effectively, with the rate constant $k_1$, there take place transitions of the localized virtual photons into the state II of the EM vacuum in the vicinity of the nucleus. We assume that it is through state II that a direct coupling of the nucleus to the EM vacuum occurs: from this state, the photons emerge into the EM vacuum, and reverse transitions of virtual photons to states I take place with the rate constant $k_2$. Let us also introduce a dynamic variable $\xi(u)$, which characterizes the level of population of the state I by the localized virtual photons, which determines the level of the notional "lubrication" needed for the movement of the particle through the EM vacuum with the velocity $u$. We will assume that, with the increasing of $u$ to the relativistic values, virtual photons localized in the domain of Casimir polarization, frontal with respect to the displacement of the particle, and also in the domain opposite to it, will be more likely to pass from state I to state II. This will correspond to an increase of the rate constant $k_1 = k_1(u)$ as $u \to c$, which should lead to a decrease in the level $\xi(u)$ of "lubrication" (the particle is partially "get stripped"). When the virtual photons in the frontal area of polarization are absent, the particle in the EM vacuum can not move. In fact, the EM vacuum acts as a "reins", holding a particle tending to escape from its polarizing shell. Thus, as the velocity $u$ increases, the potential energy of such a system increases.

Taking into account what is said, the corresponding balance equation for the variable $\xi(u)$ in the stationary case of moving a particle in the EM vacuum with the velocity $u$ can be represented as follows:

$$\frac{d\xi(u)}{dt} = -k_1 \xi(u) + k_2(1-\xi) = 0, \qquad (21)$$

so that

$$\xi(u) = \frac{k_2}{k_1 + k_2}. \qquad (22)$$

We assume, in accordance with the idea of J.J. Thompson, that the frontal and opposite domains of EM vacuum Casimir polarization in the vicinity of a particle moving with the velocity $u$ is transformed for $u \to c$ from a spherical shape to spheroid one, the surface of an ellipsoid of revolution, whose minor semi-axis $b$ is oriented in the direction of the particle velocity, whereas the major semi-axis of this ellipse remains equal to the radius $a$ of the spherical region of polarization in the case of a particle at rest or moving with non-relativistic velocities. The deformation of the domain of EM vacuum polarization as $u \to c$ can be naturally characterized by the ratio $b/a$ of the diminishing small semi-axis $b$ to the large semi-axis $a$, which is equal to $b/a = \sqrt{1-e^2}$, where $e$ is the eccentricity of the ellipse, defined as the ratio of the distance from its center to each focus to half of the major axis. It is the dependence $e = e(u)$ on the velocity $u$ that can be regarded as an indicator of increasing of the level of "nakedness" of the particle and the disappearance of "lubrication" due to the loss of localized photons in the region of EM polarization for $u \to c$. Clearly, $e \to 1$, when $b \to 0$, and the "nakedness" of the particle increases to its maximum. Furthermore, the value

$$\eta = 1/(b/a) = (1-e^2)^{-1/2} \qquad (19a)$$

can be regarded as a factor which characterizes the rate constant of losing localized photons which provide "lubrication" for moving of the particle in the EM vacuum.

With the introduction of the phenomenological relation $e(u) = u/c$ for the eccentricity of the ellipsoid of rotation, whose form is assumed by the polarization domain of the EM vacuum in the vicinity of the particle moving with the relativistic velocity $u$ relative to the basic inertial reference system, it follows from comparing (19) and (19a): $\eta = \eta_u$. If one is guided by the relations of the STR and the results of the relevant experimental studies, then one should take $k_1 = k_{10}\eta_u$, where $k_{10} \equiv k_1(0)$ and $\eta_u$ is the Heaviside factor (11). In addition, the rate constant $k_2$ should not depend on $u$. Then expression (14) can be rewritten as:

$$\xi(u) = \frac{k_2 \eta_u^{-1}}{k_{10} + k_2 \eta_u^{-1}} = \frac{k_2 \sqrt{1-u^2/c^2}}{k_{10} + k_2 \sqrt{1-u^2/c^2}}, \qquad (22a)$$

so that

$$\xi(0) = \frac{k_2}{k_{10} + k_2}, \quad \xi(u) \xrightarrow[u \to c]{} \frac{k_2}{k_{10}} \sqrt{1 - u^2/c^2}.$$

It is with the decrease of the value $\eta_u^{-1}$ with the increasing of the particle velocity and the increasing of the potential energy of the system under the disappearance of "lubrication" needed to move the body in the basic medium that is natural to connect the nature of the relativistic increase in the inertial mass and the inability to move the object in the medium with the speed of light, according to the understanding of the relation (19) by Feynman [46]. This conclusion is fully consistent with the idea of J.J. Thomson that the arising in the movement of the charged particle kinetic perturbations of the surrounding medium turn out to be equivalent to the potential (not kinetic!) energy of the moving particle, causing an increase of exactly this component of energy [39].

On the basis of the concept of a change in the geometry of the Casimir polarization domain in the vicinity of relativistic particles, upon one more relativistic effect, namely the increase of the lifetime of decaying relativistic particles can also be understood. We will cite as an example the well-studied decay process $\mu^- \to e^- + \tilde{\nu}_e + \nu_\mu$ in a ring accelerator [47]. The probability of this process is described with a high degree of accuracy by the following relation of the STR:

$$w = w_0 \eta_u^{-1} = w_0 \cdot \sqrt{1 - \frac{u^2}{c^2}}, \qquad (23)$$

where $w_0$ is the decay probability of the particle at rest. It is natural for one to attribute the lowering of the decay probability, because of the parameter $\eta_u$ decreasing with increasing velocity $u$ of the particle, to the kinetic difficulties involved in the cardinal alteration of the polarization region of the EM vacuum surrounding the particle prior to its decay. Indeed, such decay becomes possible, provided that quite definite EM vacuum polarization zones are produced in the neighborhood of the electron and electron antineutrino, as well as the muon neutrino, being formed, the particle scattering directions being governed by the laws of conservation of energy and momentum.

While the relativistic mass and lifetime increase effects are real phenomena associated with the polarization dynamics of vacuum in the vicinity of relativistic particles, the Lorentz length contraction effect [38] turns out to be associated not with the alterations occurring in the vacuum polarization region in the vicinity of relativistic particles, but with the finiteness of the rate of acquisition of data on successive measurements of the position of an object moving with a relativistic velocity.

Introduction of ideas about the EM vacuum Casimir polarization in the vicinity of nuclei and electrons made it possible to understand the essence of inertia which is one of the most mysterious phenomena of the Universe dynamics. We will define inertia as a property of a particle to move uniformly and rectilinearly with a given velocity vector with respect to the basic inertial frame of reference in the absence of any external influences on this particle. It should be recalled here that we connect the basic inertial system with the EM vacuum, all directions in which are equally probable (isotropy of space). The main question that arises in this case is to establish the cause of the apparent absence of dissipation in such a motion.

If the object is at rest with respect to the base environment – the EM vacuum, to which it is impossible to "tie in" in principle, dissipation is, by definition, excluded. As indicated above, for an inertial frame of reference with a high degree of accuracy, one can adopt a system associated with the International Celestial Reference System or another system moving with respect to this system uniformly and rectilinearly with nonrelativistic velocity. As noted above, the manifestation of dissipation in the processes under consideration should be related to the transition probabilities of virtual photons, which in this case are determined by the kinetic rate constants $k_1$ and $k_2$. With the dependence $k_1(u) = k_{10}\eta_u$ introduced above in the nonrelativistic limit, when $u \ll c$ и $k_1(u) \approx k_{10}\left(1 + u^2/2c^2\right)$, the contribution of the dissipation processes is apparently extremely small because of the limiting smallness of the term $u^2/2c^2$ in comparison with unity, that is, due to the practical indistinguishability of the quantities $k_1(u)$ и $k_{10}$. Apparently, it is in this sense that we should understand the essence of the experimentally established law of inertia for the motion of a particle, which is not acted upon by any forces, uniformly and rectilinearly relative to the inertial frame of reference. The situation can be changed when studying the motion of particles with relativistic velocities. But only experimental studies, taking into account all possible energy losses, could establish how much the dissipation effects during the motion of a relativistic particle in the EM vacuum can be significant, and also how much such dissipative losses can be compensated for by the EM vacuum energy so that motion can be considered as stationary.

In connection with the conclusion made, it should be noted that the developed approach to understanding the essence of the phenomenon of inertia when considering the uniform and rectilinear motion of bodies relative to the reference frame associated with the EM vacuum is close to the traditional view going back to Mach. According to Mach, "when we say that a certain body keeps the speed and direction of its movement in space unchanged, we imply that this ... refers to the entire Universe" ([48] 275 p.). And this Mach idea is key one, although Mach meant the static Universe and considered the law of inertia as the law of motion relative to the fixed stars (before the discovery of Hubble it was even more than 40 years). It was Mach who first introduced the notion of a basic inertial system in which "a material particle does not move in unaccelerated motion relatively relatively to space, but relatively to the center of all the other masses in the Universe" ([49] 60 p). The approach developed in this article (see also [13]) actually revives the basic ideas of Mach. This was manifested in the understanding of the essence of the gravitational interaction presented in Section 3 at the phenomenological level, taking into account the impact of all the masses of the Universe on the interaction of two attracting masses by introducing the "Mach mass". Taking into account all these factors, the basic inertial system associated with the EM vacuum was defined as the "Mach system". Therefore, it is difficult to agree with the opinion of A. Pais ([48] 280 p.) that "to this day Mach's principle has not allowed any progress in physical science; ... the problem of the origin of inertia was and remains the darkest question in the theory of particles and fields".

6. COMPARISON WITH THE EXPERIMENT

In connection with the proposed phenomenological approach to gravity, natural questions arise about the possibility to understand, based on this, the processes and phenomena that are traditionally discussed in the framework of general relativity. As is known, the main experimental confirmations of general relativity are usually associated with the experimentally registered effects: the gravitational redshift defined as the shift of the spectral lines of the source of electromagnetic waves into the long-wavelength ("red") domain of the spectrum while moving away from a massive objects [7]; the curvature of the trajectory of the light beam in the gravitational field of the Sun; an additional contribution to the secular rotation of Mercury's orbit (the "precession of the perihelion") in the direction of rotation of the planet along its orbit ([6], pp. 97-103).

All these effects, according to general relativity, are the result of gravitational time dilation, which is manifested in a slower clock rate near a large gravitating mass $M$ with the speed of light being constant in all, including non-inertial reference frames. The considered time dilation is expressed through a dimensionless parameter $\chi$, which is represented in the following form taking into account the parameters $a_{Pl}$, $a_Q$, $m_{Pl}$, $m_Q$ and $\eta_g$ introduced above:

$$\chi = \gamma \frac{GM}{c^2 R} = \frac{1}{2} \gamma \frac{M}{m_{Pl}} \frac{a_{Pl}}{R} = \frac{\gamma}{\eta_g} \frac{M}{m_Q} \frac{a_Q}{R}. \qquad (24)$$

Here, $R$ and $\gamma$ are, respectively, the characteristic linear dimension and a numerical factor for each of the considered effects. For example, for the gravitational redshift $\chi_{rs}$, measured by a distant observer at the distances much greater than the radial distance $R$ between the source of electromagnetic waves and the center of a massive body, the value of $\gamma_{rs} = 1$ [7]. For the deflection angle $\chi_S$ of the light beam in the gravitational field of the Sun, when the parameter $R$ is taken to be the Sun radius, the value of $\gamma_S = 4$ [6]. When calculating the angle $\chi_{pM}$ of displacement of the perihelion of the elliptic Mercury's orbit during one period of revolution around the Sun, the characteristic linear dimension $R$ is the quantity $a(1 - e^2)$, where $a$ is major semi-axis of the ellipse and $e$ is the eccentricity of the orbit, and then the parameter $\gamma_{pM} = 6\pi$ [6].

Expression (24), which is the ratio of the potential energy of interaction between the gravitating mass $M$ and the mass $m_0$, which is at the distance $R$, to the binding energy of this mass $m_0$ with the EM vacuum, equal to $\overline{E}_0 = m_0 c^2$, is usually considered as a relativistic factor. However, as it follows from the representation of this relation using the parameters $a_Q$ and $m_Q$, in this case one should speak not about relativistic but a quantum-mechanical nature of this factor, in accordance with the nature of the Newtonian gravity phenomenon itself. In the framework of the developed in this paper phenomenological concepts of EM vacuum as a material medium with which a basic inertial system is associated with a universal, global time for the entire Universe, it is natural to consider uniform passage of time registered by atomic standards for measuring time and frequency away from large masses and assume that the speed of light changes in the vicinity of gravitating masses. It may be reminded that initially A. Einstein related the curvature of the light beam passing near the Sun precisely with the change in the speed of light [50], as it is the case with the refraction of light rays in a medium due to changes in their propagation velocity. We are not discussing here the well-known reasons due to which this initial point of view was changed because we are also interested in a new understanding of the nature of gravity in the framework of the developed phenomenology. Therefore, referring to the relation (24), we discuss, first of all, the issues relating to the identification of the causes of the possible impact of the gravitating mass to the slowing of the speed of the passing flows of electromagnetic radiation.

It should be pointed out that, in the framework of developed ideas, the state of polarization of EM vacuum in the vicinity of each particle, if we bear in mind the population of the polarization region in the vicinity of this mass by the virtual EM quanta of different energies, depends on the spatial configuration of the polarization fields of all particles of the considered system due to the long-range character of gravitational interaction. Each of the "elementary displacements" of a single particle in the evolution of the entire system of attracting particles, in accordance with the principle of least action [51], must be accompanied by redistribution of the system of the virtual EM quanta in the vicinity of each mass with "condensation" of the shifted quanta at the corresponding excitation levels.

It is clear that the greater the mass $m_i$, the larger number of virtual EM quanta can localize in the polarization region of EM vacuum in the vicinity of this mass. Thus, in EM vacuum, which may be assumed to be "spilled" on average uniformly throughout the Universe, there are formed in the neighborhood of each mass $m_i$ the regions of "local heterogeneity". At the given length $\lambda$ of the wave of a virtual photon, its frequency $\omega_{eff} = 2\pi u_{eff}/\lambda$ is determined by the effective light velocity $u_{eff}$ in the movement of such perturbations, as it will be demonstrated below, along the interconnected, "external" for the interacting masses regions of EM vacuum. Also, the cases are possible of complete localization of the virtual photons in the near-surface regions of the condensed objects, in particular, in the propagation of the light waves in a medium with a large spatial heterogeneity [52], when $u_{eff} \to 0$, so that $\omega_{eff} \to 0$. Since virtual photons are not available for direct observation, the conclusion about the transition of a portion of the real photons into the virtual ones (this is a kind of effect, inverse to the "dynamic Casimir effect" [53, 54]) is usually associated with a decrease of the traditional scattering cross-section [55, 56].

An increase in residence time of photons in the vicinity of a gravitating massive body with a constant photon flux constitutes a region of increased stationary population of virtual photons and this manifests in slowing the stationary velocity of light near a massive body with corresponding curvature of the light beam trajectory. In fact, we are talking about an indirect impact of the mass of the Sun on the light flow, which interacts with the plasma flows and particles of different nature in various layers of the solar atmosphere, with the localization of virtual photons. As a result of this, there takes place slowing of the speed of passage of the light flow in such regions, which increases due to the growth of gravity and the density of the solar atmosphere while approaching of the passing light to the edge of the solar disk. If we have in mind the bending of the light beam emanating from a particular star near the disk of the Sun during a total solar eclipse, then as the Sun is shifting relative to the stars along the ecliptic and the curvature angle of the beam is decreasing, the light flow intensity from this star should increase due to reducing the proportion of the localized virtual photons, since the beam comes to the areas with lower gravitational influences from the Sun. Similarly, involving the concepts of slowed speed of the light flow in the scattering of light by the molecules of the medium due to increasing of the share of the formed virtual photons in the localization of the light source near the gravitating mass, the causes of the gravitational redshift can be qualitatively considered.

As for the use of relation (24) for the considering an additional contribution to the secular shift of the perihelion of Mercury's orbit, equal to just 42.98 seconds of arc per century which is 1/130 (0.77%) of the total precession speed, the situation may be more difficult. First of all, when considering the general problem of calculating the perihelion precession of the orbit for each of the solar system planets, one has to keep in mind the use of the most general expression (13*c*) for the potential energy of the paired interaction of all major masses of the solar system that may contribute to the calculated precession value of the perihelion of the planet orbit. This circumstance should manifest to the greatest extent for Mercury as the nearest planet to the Sun, because the deviation of the center of mass of the Sun from the barycenter, the center of mass of the solar system, can be up to 2% of the distance from the Sun to Mercury in the perihelion. And we must bear in mind that due to the specificities of the Sun's center of mass motion relative to the barycenter [57, 58], the value of the calculated precession of the perihelion of Mercury's orbit must change from century to century.

As for the consideration of the effects of light flows on the orbital motion of the planet, in this case it does not suffice to discuss the effects of changing of the speed of the light flows from the Sun due to the scattering processes in the environment of Mercury, the low energy plasma in the magnetosphere from the night side of the planet, the particles of a comet-like tail detected near Mercury [59]. Possibly, the decisive role can play the effects of slowing light speed in the direct impact of the flows of intense solar radiation on the nearest planet to the Sun.

From this point of view, the analysis of recent experimental results of [60] demonstrating the effect of "mass attraction by the light flow" may be of a great interest. In these experiments, the light flux generated by fluorescent lamps of various intensity, was directed above or below the mass, capable of virtually free displacements and, at the same time, connected with a sensitive device measuring the weight of this mass. In a series of experiments [61] this device unambiguously detected an increase the weight of the initial mass of 200 g, when the light flux was directed below the mass and a decrease of the weight of the mass, when the light flux was directed above the mass. The observed increase or decrease of the weight (up to thousandths of a Newton) occurred with the characteristic times of the order of tens of minutes. During the same characteristic times, there took place the relaxation changes of the weight after turning off the light fluxes. At that, the recorded maximum changes of weight increased with the increasing of the light flux intensity.

The question about the mechanism of this effect is left open in the paper [60]. In accordance with the previous consideration, it can be assumed that in the directing of the light flux of varying intensity above or below a solid sample, there is realized the effect of localization of a part of photons of the light flux, flowing around the surface, with the transformation of such photons into virtual photons, localized in the regions of Casimir polarization of the surface atoms. The unambiguous recording of the increased weight of the initial sample, when the light flux was directed below it, and the sample weight reduction when the light flux was directed above it, could mean, within the assumed mechanism, that the localization of virtual photons in the near-surface regions of the sample increases the mass of these regions. In such cases, the center of mass of the entire sample somewhat moves in the direction of the light flux in both considered cases, which, in the conditions of the experiment [60] with setting a special balance as a sensitive instrument that measures the weight of this mass, is perceived as the attraction of the mass to the light flux. Hypothetically, the additional weight $\Delta m_\omega$ associated with one virtual photon with the wave number $k = 2\pi/\lambda$ and the frequency $\omega_{eff}$, can be represented as $\Delta m_\lambda = \hbar\left(2\pi c \lambda^{-1} - \omega_{eff}\right)/c^2$. In this case, under complete localization of virtual photons in the system of surface roughness of the irradiated mass during irradiation of the sample surface by the light with the wavelength $\lambda$, we will obtain the following estimate for the total number $N_\lambda$ of the appeared localized photons:

$$N_\lambda = \frac{\lambda c}{2\pi \hbar g} \Delta P_\lambda. \qquad (25)$$

Here $\Delta P_\lambda$ is the weight change of the sample under the impact on the surface by the light flux with the wavelength $\lambda$, $g$ is the acceleration of gravity. For the values $\Delta P_\lambda = 10^{-4}$-$10^{-3}$ N, recorded in the experiment [60] for the case $\lambda = 580$ nm (yellow line of sodium), the estimate yields $N_\lambda \approx 10^{30}$-$10^{31}$.

Within the framework of the ideas about localization of a part of photons of the light flux as they pass through the medium created by the chaotically fragmented on the nano- and micrometer scales texture of the surface of a solid sample, one can also qualitatively understood other experimental regularities observed in [60]. Thus, an increase in the light flux intensity leads to increasing the observed effect of the weight change of the sample, because in this case there increases the portion of the light quanta in the laser flow turning into the virtual EM quanta localized in the vicinity of the surface atoms. Furthermore, sufficiently long (of the order of several dozens of minutes) times of establishing the value of the sample weight change under the turning on of the laser source as well as the time of the relaxation establishing of the initial sample weight after turning off the laser source, is naturally associated with the overall duration of the processes of formation of localized photons, as well as the processes of relaxation destruction of the localized photons after cessation of irradiation.

Clearly, the validity of realization of this mechanism of "attraction can" be checked by running a series of experiments by the method of [60], with variation of the parameters of roughness of both surfaces of the test sample, along which the laser beam is launched, and by changing the nature of the material from which the sample is prepared. The identification of the role of the latter factor in the formation of polarization region of the EM vacuum around the mass formed of different materials, as well as the establishment of the kinetics of the visible changes of the sample weight, is essential for the understanding of gravity as a phenomenon. In addition, we can talk about identification of quantitative relations between the changes of the "weight" of the sample and the set of 3D parameters of both surfaces [61-63] of the test sample, identified in the analysis of images obtained by atomic force microscopy. At that, the most interesting may be the so-called the spikiness parameter, characterizing the presence of the surface fragments with the most sharply varying on the submicron scale texture, since on these fragments the scattering and re-scattering of photons are most effectively realized, and, hence, the localized photons are formed. Of great interest might be the additional studies of the scattering of the laser light flux on the surface roughness, resulting in a decrease in the scattering intensity, which could help to establish the optimal conditions for the transition of real light quanta photons into the localized ones.

Thus, an increase in the light flux intensity leads to increasing the observed effect of the weight change of the sample, because in this case there increases the portion of the light quanta in the laser flow turning into the virtual EM quanta localized in the vicinity of the surface atoms. Furthermore, sufficiently long (of the order of several dozens of minutes) times of establishing the value of the sample weight change under the turning on of the laser source as well as the time of the relaxation establishing of the initial sample weight after turning off the laser source, is naturally associated with the overall duration of the processes of formation of localized photons, as well as the processes of relaxation destruction of the localized photons after cessation of irradiation.

A more complete understanding of the totality of phenomena that are realized under the influence of light flows on the real solid surfaces will allow more adequately assessing how the "direct" effects of the sun flows on the surface of Mercury in the moments of closest approach of Mercury and the Sun can contribute to the studied anomaly, the secular shift of the perihelion of Mercury's orbit. Rather, the photon localization effects are unlikely to make a significant contribution to the Mercury's mass and affect the perihelion. So the final conclusion about the contribution of various factors to the observed perihelion precession of Mercury's orbit can be made after carrying out the above calculations using more general expression (15c) for the Universal Gravitation law. So the final conclusion on the contributions of different factors to the observed precession of the perihelion of Mercury's orbit can be done only after the appropriate calculations using the general expression (13c) for the Universal Gravitation law in the inertial frame of reference associated with the center of mass of the solar system. In this case, $\vec{\xi} = \vec{R}_b$, where $\vec{R}_b$ is the radius vector of the barycenter:

$$U(\vec{\xi};\vec{r}_1,\vec{r}_2) = -Gm_{12}\left[\frac{m_1}{\left|\vec{\rho}+\frac{m_1}{m_{12}}(\vec{R}-\vec{R}_b)\right|} + \frac{m_2}{\left|\vec{\rho}-\frac{m_2}{m_{12}}(\vec{R}-\vec{R}_b)\right|}\right], \quad (13d)$$

Note also the foregoing phenomenological approach open the possibility to understand the phenomenon of change of the gravitational acceleration during the total solar eclipse (Allais effect [64]). Indeed, according to the developed ideas about the nature of gravity as a consequence of EM vacuum Casimir polarization in the vicinity of any material objects, absolutely all material bodies of the Universe are pairwise connected through the EM vacuum. Gravitational interaction of any pair of bodies is weakened if there appears between these bodies a third body which partially or completely screens these bodies from each other ("gravitational shielding") [65]. Analysis of the Allais phenomenon also allow to understand possible reasons for the observed differences in the recorded values of the gravitational acceleration value during the observation of solar eclipses in different parts of the Earth, linking the observation results to the differences in the geological features of diverse regions.

7. CONCLUDING REMARKS

The main objective of this article was to illustrate the capabilities of realistic phenomenology in establishing the essence of gravitation on the basis of introduction of the idea of EM vacuum as the basic medium of the Universe and EM vacuum "Casimir" polarization in the vicinity of material bodies. The latter idea (in Plato's understanding of the image of "*Eidos*"), supplemented by the Mach's idea of gravitational attraction of all masses of the Universe, is a kind of "thing in itself", transcendental in nature, excluding any direct experimental verification, at least in the short term. It is on this basis that it became possible to substantiate not only the Newton's formula for the Law of Universal Gravitation, but also to understand the physical cause of the unique smallness of the gravitational constant *G*. It is no doubt that in the resolution of these problems there was manifested also the gnoseological role of the relations for three Planck numbers with the simultaneous awareness of their limitations for solving the problems of gravity. The introduction of yet another, fourth arithmetical relation, a modified Weinberg's relation, allowed moving the collection of all four arithmetical relations to the level of phenomenological relations with the establishment of the physical meaning of the introduced parameters (see also [2]). This became the basis for the understanding of sufficiency of the introduction, besides $\hbar$ and $c$, of another two universal constants $m_Q$ and $R_H$, in order to present at the phenomenological level the main fundamental interconnections in the Universe and to focus on the development of the general theory of gravity and the dynamics of the Universe within not the $\hbar G c$-, but the $\hbar m_Q R_H c$-plan. It is necessary to emphasize that the main essence of the $\hbar m_Q R_H c$-plan under consideration is the postulated dependence of the fundamental parameters $\hbar$, $c$ и $m_Q$ from the global time of the Universe. Thus, in accordance with the results of [2] and the relation (5), is postulated the change in time value of the gravitational constant $G$. Last conclusion is in line with the results of [66].

The presented phenomenology and the introduction of EM vacuum as the basic medium, a kind of "ether" of the Universe, allows coming closer, as noted above, to the understanding of the physical nature of some *a priori* introduced images in the special theory of relativity (the reason of limitation of the maximum speed of movement of the material body by the speed of light $c$ in vacuum; genesis of the relation $\overline{E}_0 = m_0 c^2$) and quantum mechanics (the essence of the dual image of the "wave-particle", the impossibility of simultaneous determination of the spatial position of the particle and its momentum).